\documentclass[twoside]{article}
\usepackage{curves,fleqn,espcrc2}
\usepackage{amssymb}
\usepackage{graphicx}

\newcommand{\AmS}{{\protect\the\textfont2
  A\kern-.1667em\lower.5ex\hbox{M}\kern-.125emS}}

\title{String breaking mechanisms induced 
by magnetic and electric condensates }

\author{Ferdinando Gliozzi and Antonio Rago \address{Dipartimento di Fisica 
       Teorica
        dell'Universit\`a di Torino, and \\
       Istituto Nazionale di Fisica Nucleare, via P.Giuria 1, 
       I-10125 Torino,Italy} }      
\begin{document}
\def\de{\partial}
\def\oh{\frac{1}{2}}
\newcommand\hb{\hbar}
\newcommand{\bra}{\langle}
\newcommand{\ket}{\rangle}
\newcommand{\um}{\frac 12}
\newcommand{\eq}{\begin{equation}}
\newcommand{\en}{\end{equation}}

\begin{abstract}
The normal confining phase of gauge theories is characterised by
the condensation of magnetic monopoles and center vortices. Sometimes
in coupled gauge system one finds another phase with simultaneous
condensation of electric and magnetic charges. In both phases
 the confining string breaks down at a given scale because of pair creation,
 however the mechanism is different. In the former case the string breaking is
 a mixing phenomenon  which is invisible  in the Wilson loop. On the
 contrary, in presence of both electric and magnetic 
condensates the string breaking can be observed even in the Wilson loops. 
Numerical experiments on a 3D $Z_2$ gauge-Higgs system neatly show
this new phenomenon. 

\end{abstract}
\vspace{1pc}

\maketitle
\section{INTRODUCTION}

Center vortices \cite{th} and magnetic monopoles \cite{tm} are widely
believed to be the most important degrees of freedom for confinement 
in Yang Mills theories. Plausibility arguments suggest that  
 large center vortices or  magnetic condensates imply area law for
 the Wilson loop. 

It is worth noting that while these arguments are generally applied
 to pure Yang Mills models, they hold true also 
in gauge theories coupled to matter. This can explain a surprising
phenomenon observed in all these kind of coupled systems: although 
the potential between static sources flattens at large distances because
of the screening produced by pair creation, this flattening (called
string breaking) is invisible in the Wilson loop: it continues to obey
an area law   in full QCD \cite{qcd}  even at
distances where the static charges are completely screened.
The point is that in gauge theories coupled to matter the basis of the 
operators has to be enlarged \cite{cm} in order to get a reliable
estimate of the potential. In this way it has been observed the
breaking of the confining string in Higgs models \cite{ks} and in QCD 
\cite{milc}. 
So the fact that large Wilson loops obey an area law even  in
coupled systems may be considered  as a further support to the 
plausibility arguments for the confinement mechanisms.

In this work we use such a special property, which is
also suggested by the underlying string picture of  the confining phase
\cite{gp}, as a tool to test the validity of various confinement
criteria. Indeed in  coupled gauge systems there are different 
vacua, distinguished by different entities which condense.
In the simple model we use as a guide, namely the 3D $Z_2$
gauge-Higgs model, there are five kinds of vacua (see Fig.1).
A simple, rigorous, argument shows that the only confining vacua 
(i.e. those with Wilson loop obeying an area law) are those in which both
magnetic monopoles \underline{and}  center vortices condense. 
Luckily there is a non-confining vacuum with monopoles but without
vortices (region III of Fig.1), where this result has been checked. 
 
This model has two different confining vacua. Both have
center vortex and magnetic monopole condensates, but one of them has
also an electric condensate, i.e. the Higgs field has a vacuum
expectation value different from zero.

The former fulfils all the requirements of the confinement
criteria, so an area law is expected; here we  found that  the
Wilson loop obeys a perfect area law even at distances larger than
five times the string breaking scale.

The latter  can be identified with the torn phase predicted in 
Ref.\cite{gp}: the Wilson loop obeys an area law below a given
scale. Above this threshold we observed for the first time a clear
signal of string breaking (see Fig.3). A similar
phase in 4D $SU(2)$-Higgs model has been reported at this 
conference \cite{bfgo}.

\section{THE MODEL}
The action of a  3D $Z_2$ gauge theory coupled to  
 matter can be written as 
\eq
S(\beta_G,\beta_I)=-\beta_I\sum_{\langle ij\rangle}
\sigma_i U_{ij}\sigma_j-
\beta_G\sum_{plaq.}U_{\square}~,
\en
where both the link variable $U_{ij}\equiv U_\ell$  and the matter
field $\sigma_i$  take values $\pm1$ and
$U_{\square}=\prod_{\ell\in\square}U_\ell$.
In this model the construction of center vortex configurations is 
straightforward: to each frustrated plaquette
(i.e. $U_{\square}=-1$) assign a vortex in the dual link.  Since
the product of the plaquettes belonging to any elementary cube is 1,
center vortices form closed subgraphs of even coordination number.
Thus a connected vortex subgraph contributes to a given Wilson loop
 $W(C)$ only if an {\sl odd} number of lines are linked to it. 
The Wilson loop is a vortex counter modulo 2. Finite
vortex subgraphs contribute only to the perimeter term of $\langle
W(C)\rangle$. This expectation value obeys an area law only if there is
an \underline{infinite} vortex subgraph. The plausibility arguments
mentioned above would suggest the reverse statement:  large center 
vortices should imply confinement. In the following we show that 
confinement requires a further constraint, generated by Kramers Wannier
 duality. Under such a
transformation the model is self-dual and the vacuum expectation value of the
 Wilson loop coincides with the expectation value of the 't Hooft loop
 at the dual point

\eq
\langle W(C)\rangle_{\beta_G\,\beta_I}=
\langle \widetilde{W}(C)\rangle_{\tilde\beta_I\,\tilde\beta_G}~~,
\label{duality}
\en
with   $\tilde\beta=-\um\log(\tanh \beta)$ and 
\eq
\widetilde{W}(C)=\exp(-2\tilde\beta_I\sum_{
\langle ij\rangle\in\Sigma}\sigma_iU_{ij}\sigma_j)
~,~~ \partial \Sigma=C\;.
\en
Performing a suitable field transformation first introduced by Fortuin and
Kasteleyn  (FK) \cite{fk}, $\widetilde{W}(C)$ becomes a counter of a 
new kind of lattice subgraphs, namely of the FK clusters topologically
linked to it \cite{gv}. 
The  allowed FK clusters in the direct lattice are the subgraphs   
for which any closed path
within the cluster is linked to an even number of center vortices \cite{cg},
hence no $Z_2$ magnetic flux can pass through the loops of the  FK clusters
 \footnote{This is a $Z_2$ analogue of the Meissner effect.}.

We can now apply the same line of reasoning used for the center
vortices. Finite clusters  can link with the 't Hooft loop  only
along its perimeter. Therefore they contribute only to the perimeter
term. Owing to the duality relation
(\ref{duality}), the Wilson loop obeys an area law only if 
there is  an infinite, percolating,
FK cluster in the \underline{dual}  phase. An infinite FK cluster in 
the dual description corresponds exactly to the dual Higgs mechanism 
of 't Hooft and Mandelstam \cite{tm}: a disorder field $\tilde\sigma$
carrying a $Z_2$ magnetic charge acquires a non-vanishing vacuum
expectation value. Thus confinement implies both infinite center
vortex subgraph \underline{and} magnetic monopole condensation. 
In pure gauge theory these two requirements coincide, while in the 
coupled system the situation is more tricky.    
 
\vskip .6 cm
{\hskip -.3 cm\includegraphics[height=24pc,width=17pc]{phases.epsi}}

\vskip -4.7 cm
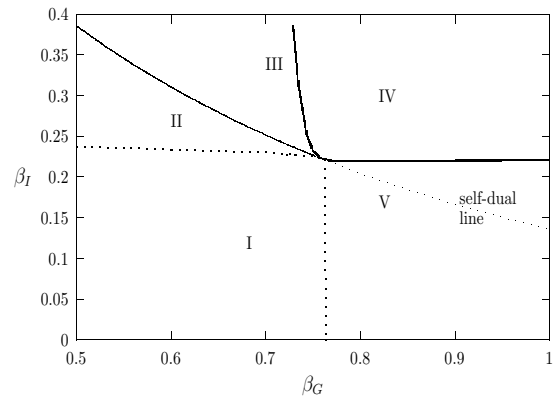
Figure 1.  Phase diagram of the $Z_2$ gauge Higgs model.
\vskip .5 cm
In the coupled theory we have two kind of interesting subgraphs:
center vortices {\sl or}  FK clusters in the \underline{dual} 
lattice describe the gauge field degrees of freedom; FK clusters in the
\underline{direct} lattice describe the charged Higgs
matter. 
There are three kind of infinite clusters: {\sl i)} FK cluster in the
dual lattice (magnetic condensate), {\sl ii)} FK cluster in the direct
lattice (electric condensate), {\sl iii)} large center vortices in the
dual lattice. The size of  these clusters in finite lattices is
proportional to the volume. Straightforward numerical experiments show that 
 they are distributed in the phase diagram \cite{js} according to
Fig.1 and Tab.1
\vskip .2 cm
 \centerline{Tab. 1}
\vskip .3 cm
\begin{tabular}{cccc}
\hline
phase& magnetic& electric& large\\
~& condensate&condensate& vortices\\
\hline
I&yes&no&yes\\
II&yes&yes&yes\\
III&yes&yes&no\\
IV&no&yes&no\\
V&no&no&no\\
\hline
\end{tabular}
\vskip .2 cm
Region III has a magnetic condensate but no large center vortices then 
there is no confinement, as confirmed by numerical tests.  The region
I and II are confining. The former is a normal confining phase
\cite{gp}: the potential extracted from an enlarged basis shows the
expected string breaking (see Fig.2), while the Wilson loop obeys a
perfect area law even at large scale. This has been  checked up to
distances of the order of five times the string breaking scale \cite{fg}. 
\vskip 1.3 cm
\includegraphics[height=13pc,width=17pc]{pot.epsi}
\vskip .4 cm
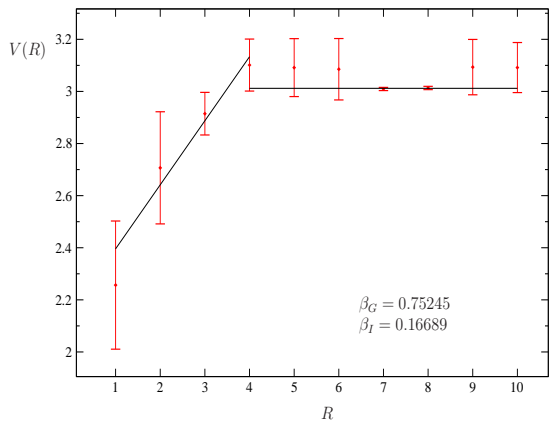
Figure 2. The static potential in the region I
\vskip .5 cm
 The latter is a torn phase produced
by the Meissner effect we alluded before: the simultaneous presence of
infinite  clusters in the direct and the dual lattices with the
constraint of no frustration induces strong correlations on vortex
lines which produce a visible string breaking effect also in the
Wilson loop, as Fig.3 clearly shows.

\includegraphics[height=13pc,width=17pc]{torn.epsi}
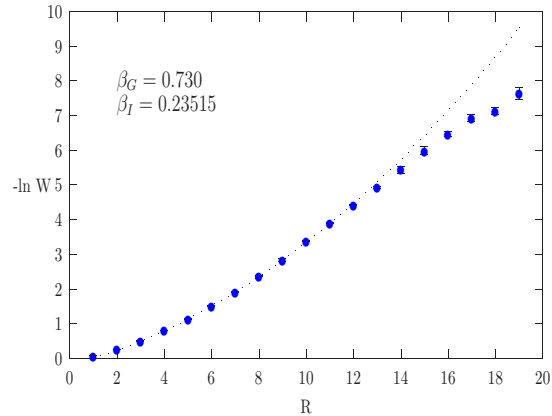
Figure 3. $-\ln W(R,R)$ as a function of  $R$ in the region II (torn phase).

\end{document}